# Bound soliton fiber laser

**D. Y. Tang, B. Zhao, D. Y. Shen and C. Lu**

School of electrical and Electronic Engineering, Nanyang Technological University, Singapore

**W. S. Man and H. Y. Tam**

Department of Electrical Engineering, Hong Kong Polytechnic University, Hung Hom, Hong Kong

Experimental study on the soliton dynamics of a passively mode locked fiber ring laser firstly revealed a state of bound soliton operation in the laser, where two solitons bind together tightly with fixed pulse separation. We further report on the properties of the bound-soliton emission of the laser. In particular, we demonstrate both experimentally and numerically that, like the single pulse soliton operation of the laser, the bound soliton emission is another intrinsic feature of the laser.

Self-started, passively mode locked fiber lasers as a potential source of ultrashort optical pulses have been intensively investigated [1-7]. A generic feature of the lasers found is that under suitable operation conditions, they can emit the so-called soliton pulses - optical pulses that are sech-form shaped and have nearly transform-limited bandwidth-duration product. Under a soliton operation not only the output pulses of the lasers become ultrashort, but also the pulse–to-pulse energy and peak power become ultra-stable, which was found to be quantum noise limited [8].

Soliton emission of the lasers is a natural consequence of the nonlinear pulse propagation in the fiber cavity, where due to the balanced action between the fiber optical Kerr effect and the cavity dispersion on a pulse, whose shape and duration become unchanged with the propagation. Although in a passively mode-locked fiber laser, apart from the optical fiber other optical components such as the gain medium and output coupler coexist in the cavity, which affect the detailed dynamics of the formed pulses. It is nevertheless demonstrated that under weak influence of them, the average dynamics of the solitons could be well described by the nonlinear Schrödinger equation [9]. In this paper we report on states of bound soliton emission and their properties in a passively mode-locked fiber ring laser. We show both experimentally and numerically that apart from the single pulse soliton emission, the laser can also emit stable, closely spaced soliton pairs. In particular, when operating in the regime, the bound soliton pair is the only stable structure of the solitary wave in the laser, it as a unit has exactly the same features as those of a single pulse soliton. For this reason, we refer the fiber laser as a bound-soliton fiber laser.

Bound states of solitons have recently been predicted in the coupled nonlinear Schrödinger equations [10], and the quintic complex Ginzburg-Landau equation [11-12]. Formation of bound solitons was explained as a result of direct soliton interaction: Solitons formed in these systems have an oscillating tail, when they interact, their effective interaction potential has spatial local minima, which give rise to stable bound solitons. Bound solitons this formed are characterized as that they have fixed, discrete pulse separations, which are independent of soliton propagation. For a passively mode-locked fiber laser when the influences of laser gain and cavity losses, and /or fiber birefringence become strong, its dynamics is actually described by the Ginzburg-Landan equation, or its coupled forms [13].

The passively mode-locked fiber ring laser used in our experiments is schematically shown in Fig. 1. It has a ring cavity of about 5.5 meters long. The cavity comprises of a 3.5-meter long 2000ppm erbium-doped fiber with a group velocity dispersion of about –10 ps/nm km, one piece of 1 meter long single mode dispersion shifted fiber, whose group velocity dispersion is –2 ps/nm km and one piece of 1 meter long standard single mode fiber (SM-28). The nonlinear polarization rotation technique [5] is used to achieve the self-started mode locking in the laser. To this end a polarization dependent isolator together with two polarization controllers, one consisting of two quarter-wave plates and the other of two quarter-wave plates and one half-wave plate, is used to adjust the polarization of light in the cavity. The polarization dependent isolator and the polarization controllers are mounted on a 7cm long fiber bench to achieve accurate polarization adjustments. The laser is pumped by a pigtailed InGaAsP semiconductor diode of wavelength 1480nm. The output of the laser is taken via a 10% fiber coupler and analyzed with an optical spectrum analyzer (Ando AQ-6315B) and a commercial optical autocorrelator (Inrad 5-14-LDA). A 50 GHz wide bandwidth sampling oscilloscope (Agilent 86100A) and a 25 GHz high-speed photo-detector (New Focus 1414FC) were used to study the evolution of the bound solitons in the cavity.

In two previous papers [14,15] we have reported on how to experimentally achieve the bound-soliton operation and some of the basic properties of the observed bound solitons, respectively. The key technique used to get stable bound soliton operation of the laser is the intra-cavity mode locking of the dynamical laser modes. With this technique the influences caused by the random environmental noise and the dispersive waves on the soliton interaction could be suppressed. In our previous experiment we have also confirmed that the observed bound solitons have the characteristics of discrete, fixed soliton separations which are independent of soliton propagation, as predicted theoretically.

The states of bound solitons are usually abruptly formed in our laser from a single pulse soliton state through adjusting the orientations of the waveplates. However, if their orientations are already appropriately set, bound soliton operation can also be directly obtained by simply increasing the pump power beyond the mode-locking threshold. For the completeness of the paper we have shown again in Fig. 2 the optical spectra of the bound solitons observed in our laser. In order to check if the formation of bound states of solitons is a generic feature of the passively mode-locked fiber lasers, in the current experiment we have deliberately changed the cavity property of our laser. Our experimental results confirmed that with exactly the same technique and procedure, the bound states of solitons could be re-obtained in the laser. With the new fiber laser setup we have obtained four discrete states of bound solitons as shown. Comparing them with those obtained with our previous fiber laser setup, except the different pulse separation and soliton pulse width, they have exactly the same properties. Namely, they are ultra stable and have discrete, fixed pulse separations that are neither dependent on the soliton propagation nor on the exact laser operation conditions, such as the pump intensity and the cavity detuning etc. The change in the soliton pulse width is a natural result of the total cavity dispersion change. In turn it causes the pulse separation change of the bound solitons. Nevertheless, once a laser setup is fixed, the pulse separations of bound solitons obtainable are fixed. Analyzing the pulse separations among the four bound soliton states, surprisingly, they have again the relationship that one is roughly twice of the other as can be easily checked from the spectral modulation of the bound solitons shown in Fig. 2, indicting that this relationship could be a universal property of the bound solitons in the fiber laser system and it is independent of the concrete laser setup.

Like the single pulse soliton operation of the laser, depending on the pumping strength, multiple bound solitons can also coexist in the laser cavity. As an example Fig. 3 shows two oscilloscope traces of such states. Fig.3a shows that twelve bound soliton pulses coexist in the laser cavity. Note that the pulse separation between the bound solitons is about 920fs and the soliton pulse width of the laser is about 325 fs when a sech-form pulse shape is assumed. As the detector can’t separate the pulse separation, in the oscilloscope traces the bound solitons looks like a single pulse. When multiple bound solitons coexist in the laser cavity and they are far apart separated, except the spectral or signal strength is increased, there is no difference observable on the optical spectrum and the measured autocorrelation traces comparing with those of single bound soliton pair in the cavity. Based on this experimental evidence together with the result shown in Fig. 3, we conclude that all the bound solitons in the cavity have exactly the same property: namely the same soliton separation and same pulse energy (there are small peak intensity fluctuations on the pulses shown in the measured oscilloscope traces. We have checked it and determined that they were due to the sampling problem of the sampling oscilloscope. By reducing the recorded time range of the trace, the fluctuation disappeared and all the pulses had exactly the same height). In all of our experiments in a stable state no bound solitons of different pulse separations or different pulse energy have been observed to coexist, and no unbound solitons with the bound solitons, or bound solitons of higher order (with more than two soliton pulses binding together) have been observed to coexist, which suggests strongly that in the parameter regime, the only stable pulse structure is the bound soliton pairs of the fixed pulse separation.

Carefully increasing or decreasing the pump power, the number of bound soliton pairs can be increased or reduced one by one. Fig. 3a and b show a case of two consecutive states when the pump power is reduced. Initially there were twelve bound solitons in the cavity, reducing the pump power to a certain value one bound soliton pair were suddenly destroyed simultaneously. After losing one pair of the bound solitons, the rest of the bound solitons rearranged their relative positions in the cavity and stabilized finally at a new pattern as shown in the Fig. 3b. In our experiments we found that the dispersive waves play an essential role in the long rang interaction among the bound solitons. Mediated through the dispersive waves, the bound solitons can interact among themselves and form various states of multiple bound soliton operation. In fact Fig. 3 shows two cases of the so-called quasi-harmonic mode-locking by some authors in the single pulse soliton operation of fiber lasers. States of harmonic mode locking of the bound solitons have also been frequently observed. Fig. 4 shows as an example one of these states. In the current case eight bound solitons equally spaced in the cavity. We have also observed situations where several tens of bound solitons coexist in the cavity and form harmonic mode locking. Depending on the interaction between the bound solitons, bunches of bound solitons have also been revealed in the laser as shown in Fig. 5. Several bound solitons can tightly couple together locally to form a bound soliton bunch, and the bunch move with the fundamental repetition rate in the cavity, exactly as also observed in the single pulse soliton case [1-2,4].

When the bound soliton pairs interact directly with each other, they behave as a unit: either colliding elastically or simply crossing each other without damaging their bindings or altering their pulse separations. With the high-speed sampling oscilloscope and detector we could monitor in the real time interact between two bound soliton pairs as shown in Fig. 6. The interaction between the bound soliton pairs clearly exhibits the particle-like nature of the soliton interaction. Bound states of the bound solitons mediated through the dispersive waves were occasionally observed. However, these bound states are unstable, a small perturbation can destroy it. All these experimental findings demonstrate that the bound soliton pairs as a unit has exactly all the same properties as those of a single pulse soliton in the laser. Based on their properties we conclude that the observed bound states of solitons could be another type of solitary wave in the laser and the bound soliton operation could be an intrinsic feature of the system.

To confirm our conclusions, we have also numerically simulated the operation of our laser. For this purpose we have built up a theoretical model based on the coupled extended nonlinear Schrödinger equations. Our model takes into account the birefringence of the optical fiber and the perturbations caused by the laser gain and cavity loss, and has a form:

$$
\begin{aligned}
\frac{\partial u}{\partial z} &= i\beta u - \delta\frac{\partial u}{\partial t} - \frac{i}{2}\kappa''\frac{\partial^2 u}{\partial t^2} + i\gamma\left(|u|^2 + \frac{2}{3}|v|^2\right)u + \frac{i\gamma}{3}v^2u^* + \frac{g}{2}u + \frac{g}{2\Omega_g^2}\frac{\partial^2 u}{\partial t^2} \\
\frac{\partial v}{\partial z} &= -i\beta v + \delta\frac{\partial v}{\partial t} - \frac{i}{2}\kappa''\frac{\partial^2 v}{\partial t^2} + i\gamma\left(|v|^2 + \frac{2}{3}|u|^2\right)v + \frac{i\gamma}{3}u^2v^* + \frac{g}{2}v + \frac{g}{2\Omega_g^2}\frac{\partial^2 v}{\partial t^2}
\end{aligned}
\qquad (1)
$$

Where *u* and *v* are the two normalized slowly varying pulse envelopes along the slow and the fast axes, respectively. $2\beta = 2\pi\Delta n/\lambda$ is the wave-number difference, $2\delta = 2\beta\lambda/2\pi c$ is the inverse group-velocity difference. κ” is the dispersion parameter, γ is the nonlinearity of the fiber. g is the laser gain coefficient and $\Omega_g$ is the bandwidth of the laser gain. Gain saturation of the laser is considered through writing

$$
g = \frac{g_0}{\left[1 + \frac{\int \left(|u|^2 + |v|^2\right)dt}{E_s}\right]} \qquad (2)
$$

where $g_0$ is the small signal gain and $E_s$ is the saturation energy. To also account the effects of other cavity components and the cavity feedback, we simulate the laser by simply letting the light circulate in the cavity, whenever it meets a cavity component we then discretely multiply the component's transformation matrix to the light. In all of our simulations where it is possible, we have used the actual value of the laser parameters.

Using the same theoretical model we have previously simulated the dynamics of single pulse soliton lasing of passively mode-locked fiber lasers, and correctly reproduced the features such as the intrinsic soliton wavelength tuning [16], soliton sideband asymmetry and subsideband generation [13]. In the current simulation we found that by appropriately choose the linear cavity transmission of the laser and the polarization orientation of the polarization dependent isolator, bound states of solitons could actually be reproduced in our model. In particular, we found that with the parameter settings, bound solitons are the only stable state. Fig. 7 shows results of our simulations. Three stable bound soliton states with different pulse separations have been revealed in our simulations under the laser parameter settings. In our simulations in order to find out the bound soliton states we simply fix the laser parameters and input either random initial noise waveform or pulses of different shapes to start our simulations, we then wait for until a stable state is achieved. We found that independent of the initial conditions the laser output always settled down to one of the bound soliton states. Similar to the experimental results, the obtained stable bound solitons exhibit discrete, fixed pulse separations. Once a state of bound solitons is obtained, it will remain there for several thousand rounds of calculation, even under slightly changed parameters such as the small signal gain and saturation energy, indicating that it is a stable state of the system. Our numerical simulations have fully confirmed our experimental observations.

In summary we have experimentally studied properties of bound soliton emission in a passively mode-locked fiber ring laser, and observed features such as the bound soliton energy quantization, quasi- and harmonic bound soliton mode locking, bound soliton bunching and particle-like interaction between bound solitons. Our experimental results have also revealed that no single pulse soliton can coexist with the bound solitons, and when multiple bound solitons coexist, all bound solitons have exactly the same properties. Exactly the same features have also been observed for the single pulse soliton in the fiber soliton lasers, which suggests strongly that, like the single pulse soliton, the observed bound solitons are another fundamental structure of stable light pulse in the laser, and the bound soliton operation of the laser is in fact an intrinsic feature of the fiber laser. Finally, we note that numerical simulations on the laser operation have also confirmed the existence of bound solitons in the laser and its property of discrete, fixed soliton separations, which further support our conclusions.

Acknowledgement: W. S. Man and H. Y. Tam acknowledge support by a university research grant of The Hong Kong Polytechnic University.

# References:

[1] D. J. Richardson, R. I. Laming, D. N. Payne, V. J. Matsas, M. W. Phillips, Electron. Lett., 27, 1451-1452 (1991).

[2] A. B. Grudinin, D. J. Richardson and D. N. Payne, Electron. Lett., 27, 1860-1861 (1993).

[3] C. J. Chen, P. K. A. Wai and C. R. Menyuk, Opt. Lett., 17, 417-419 (1992).

[4] A. B. Grudinin, D. J. Richardson and D. N. Payne, Electron. Lett., 28, 67-68 (1992).

[5] V. J. Matsas, D. J. Richardson, T. P. Newson and D. N. Payne, Opt. Lett., 18, 358-360 (1993).

[6] M. J. Guy, D. U. Noske and J. R. Taylor, Opt. Lett., 18, 1447-1449 (1993).

[7] K. Tamura, E. P. Ippen, H. A. Haus and L. E. Nelson, Opt. Lett., 18, 1080-1082 (1993).

[8] H. A. Haus and A. Mecozzi, IEEE J. Quantum Electron. 29, 983-996 (1993).

[9] S. M. J. Kelly, K. Smith, K. J. Blow and N. J. Doran, Opt. Lett., 16, 1337-1339(1991).

[10] B. A. Malomed, Phys. Rev. A., 45, R8321-R8323 (1992).

[11] B. A. Malomed, Phys. Rev. A., 44, 6954-6957 (1991).

[12] N. N. Akhmediev, A. Ankiewicz, and J. M. Soto-Crespo, Phys. Rev. Lett., 79, 4047-4051 (1997).

[13] D. Y. Tang, S. Fleming, W. S. Man, H. Y. Tam and M. S. Demokan, J. Opt. Soc. Am. B, 18, 1443-1450 (2001).

[14] D. Y. Tang, W. S. Man, H. Y. Tam and P. D. Drummond, Phys. Rev. A., 64, 033814, 2001.

[15] D. Y. Tang, B. Zhao, D. Y. Shen, W. S. Man and H. Y. Tam, Opt. Commun. 2002, submitted.

[16] W. S. Man, H. Y. Tam, M. S. Demokan, P. K. A. Wai and D. Y. Tang, J. Opt. Soc. Am. B, 17, 28-33(2000).

# Figure captions:

Figure 1: A schematic of the passively mode-locked fiber laser. λ/4: quarter-wave plate; λ/2: half-wave plate; PI: polarization-dependent isolator. WDM: wavelength-dividend-multiplexer.

Figure 2: Optical spectra of the bound states of solitons observed.

cc) With spectral modulation period of about 9.1nm.

dd) With spectral modulation period of about 4.7nm.

ee) With spectral modulation period of about 2.6nm.

ff) With spectral modulation period of about 1.3nm.

The soliton pulse width is about 325 fs and the separations between the bound solitons can be calculated from the corresponding spectral modulation periods.

Figure 3: Oscilloscope traces of multiple bound solitons in the cavity. The cavity round-trip time is about 26 ns for the laser.

a) With twelve bound solitons in the cavity.

b) With eleven bound solitons in the cavity.

From a to b the pump laser power is slightly reduced.

Figure 4: Harmonic mode locking of the bound solitons. Eight bound solitons are equally spaced in the cavity forming the special mode-locked state.

Figure 5: Bunching of the bound solitons. Seven bound solitons coupled tightly together. No relative movement among the solitons.

Figure 6: Collision between two bound soliton pairs. After collision they still remain as bound solitons of the same property.

Figure 7: States of bound solitons calculated from the theoretical model.

a, b, c) Bound solitons with different pulse separations.

e, d, f) The corresponding optical spectra.

Parameters used are $\gamma = 3W^{-1}km^{-1}$, κ”= -2 ps /nm km (for dispersion shifted fiber), κ” = – 10 ps/nm km (for erbium doped fiber), $g_0 = 254$, $\Omega_g = 2\pi\times 10$ THz, $E_s = 110$, cavity length L = 11m, beat length $L_b = L/2$.

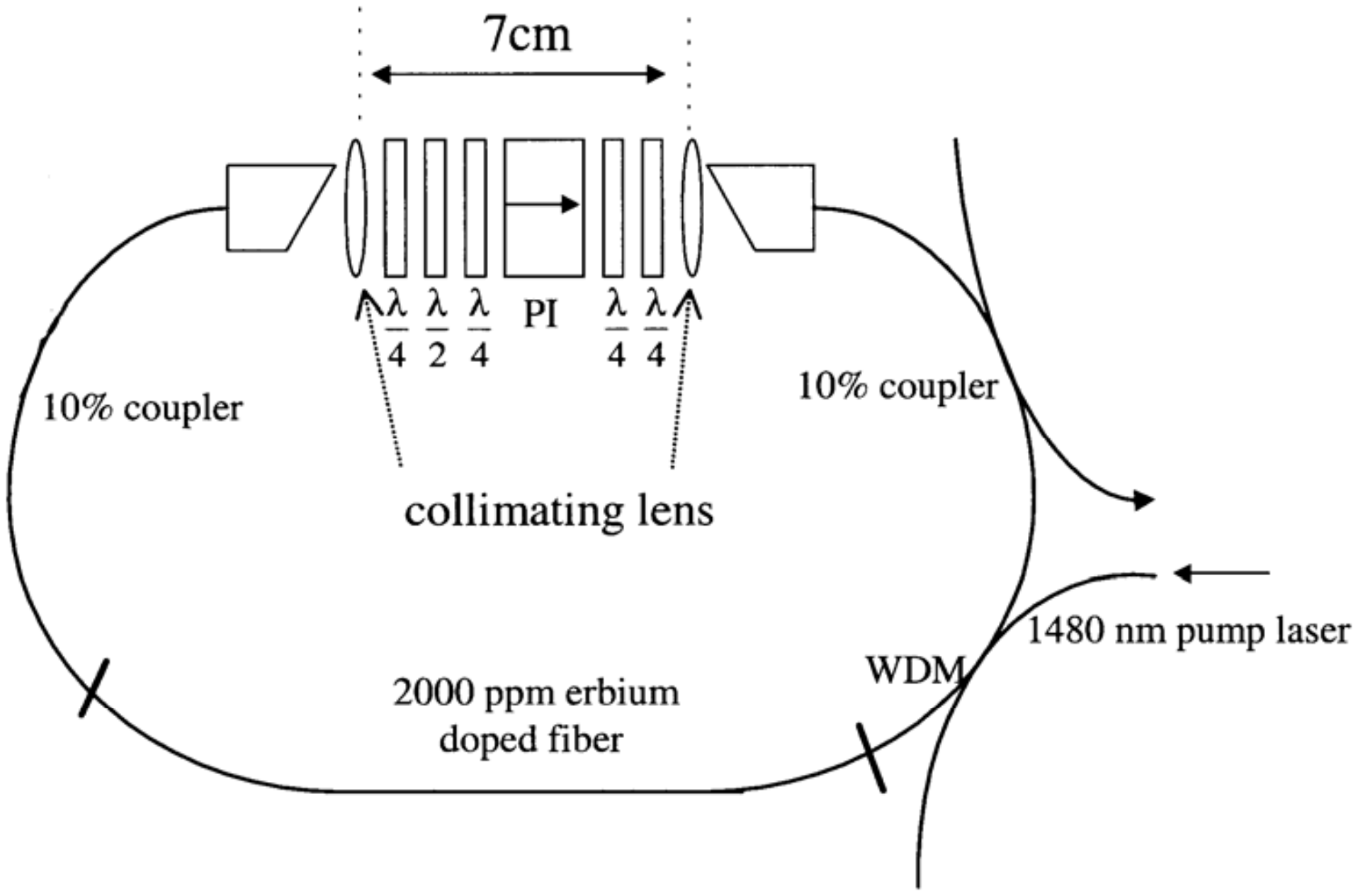

Figure 1

D. Y. Tang et al

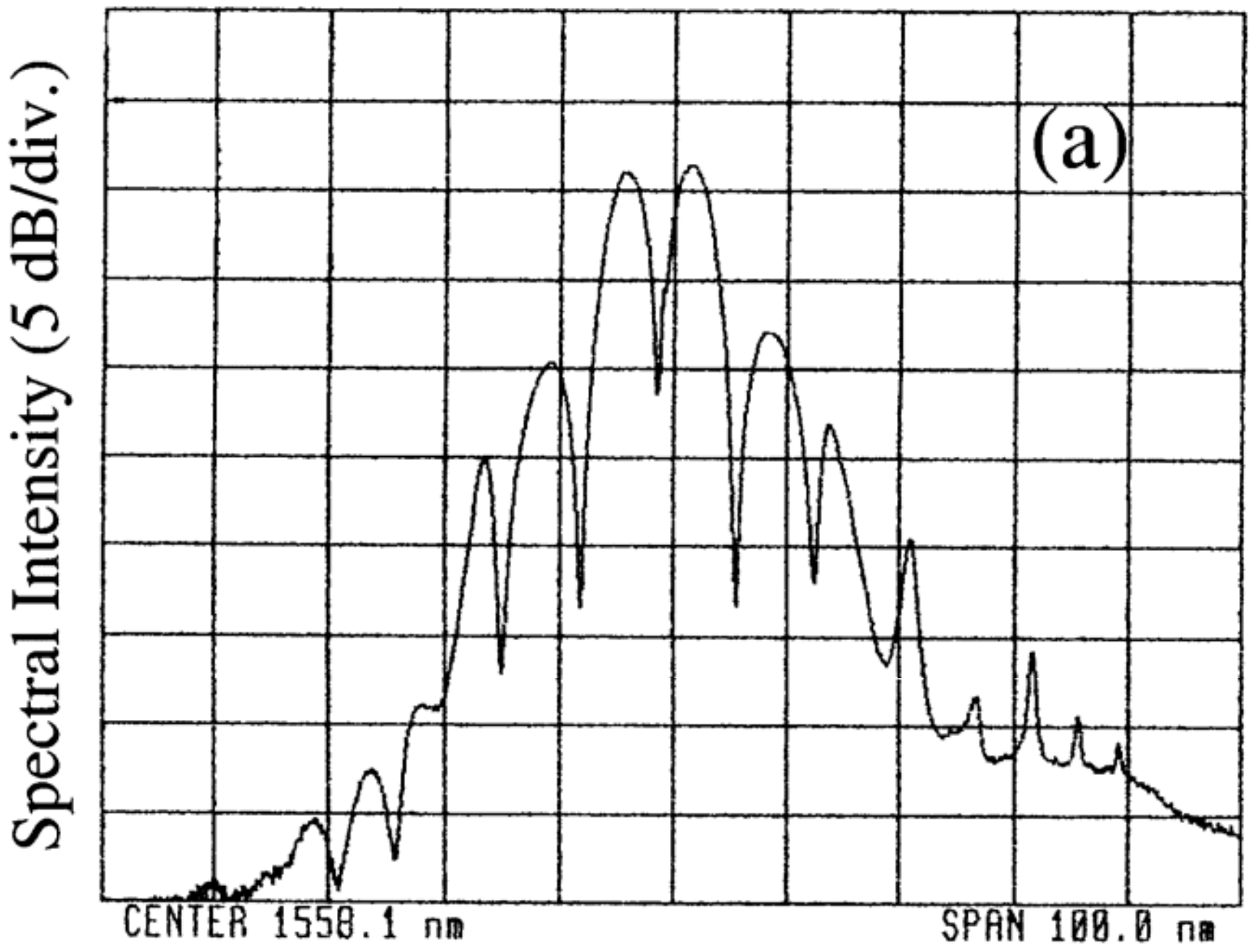

Figure 2(a)

D. Y. Tang *et. al.*

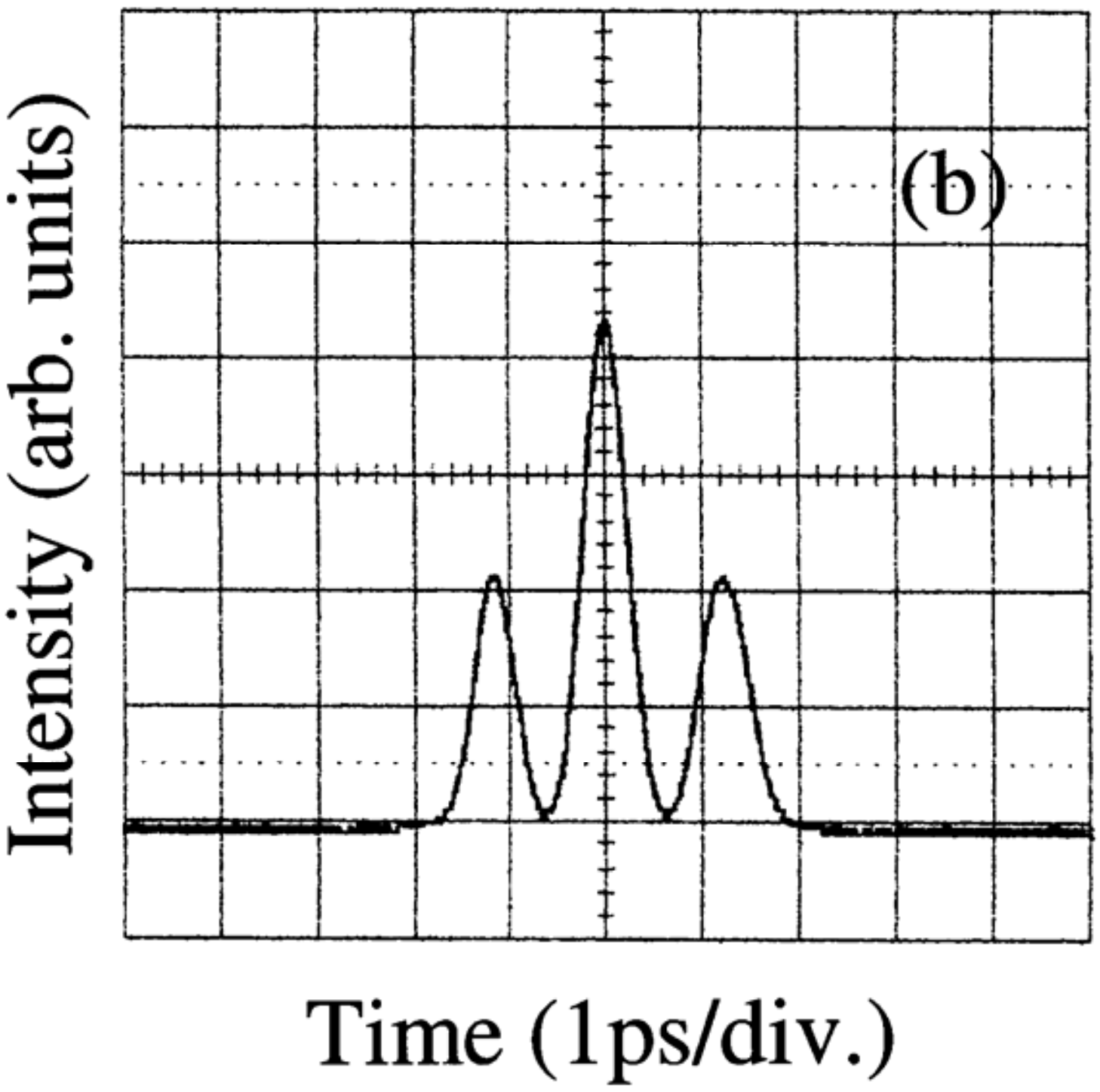

Figure 2(b)

D. Y. Tang *et. al.*

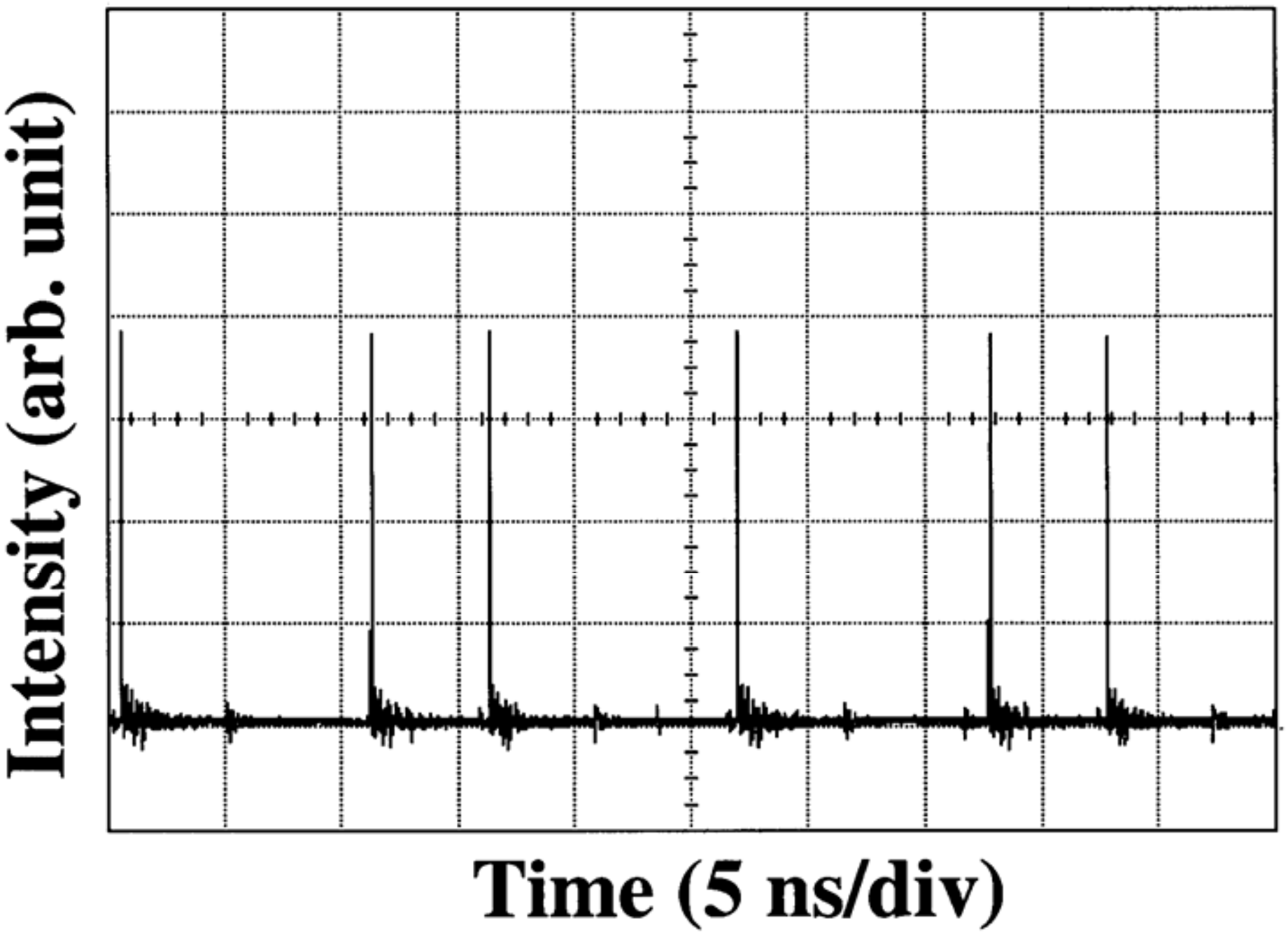

Figure 3.

D. Y. Tang *et. al.*

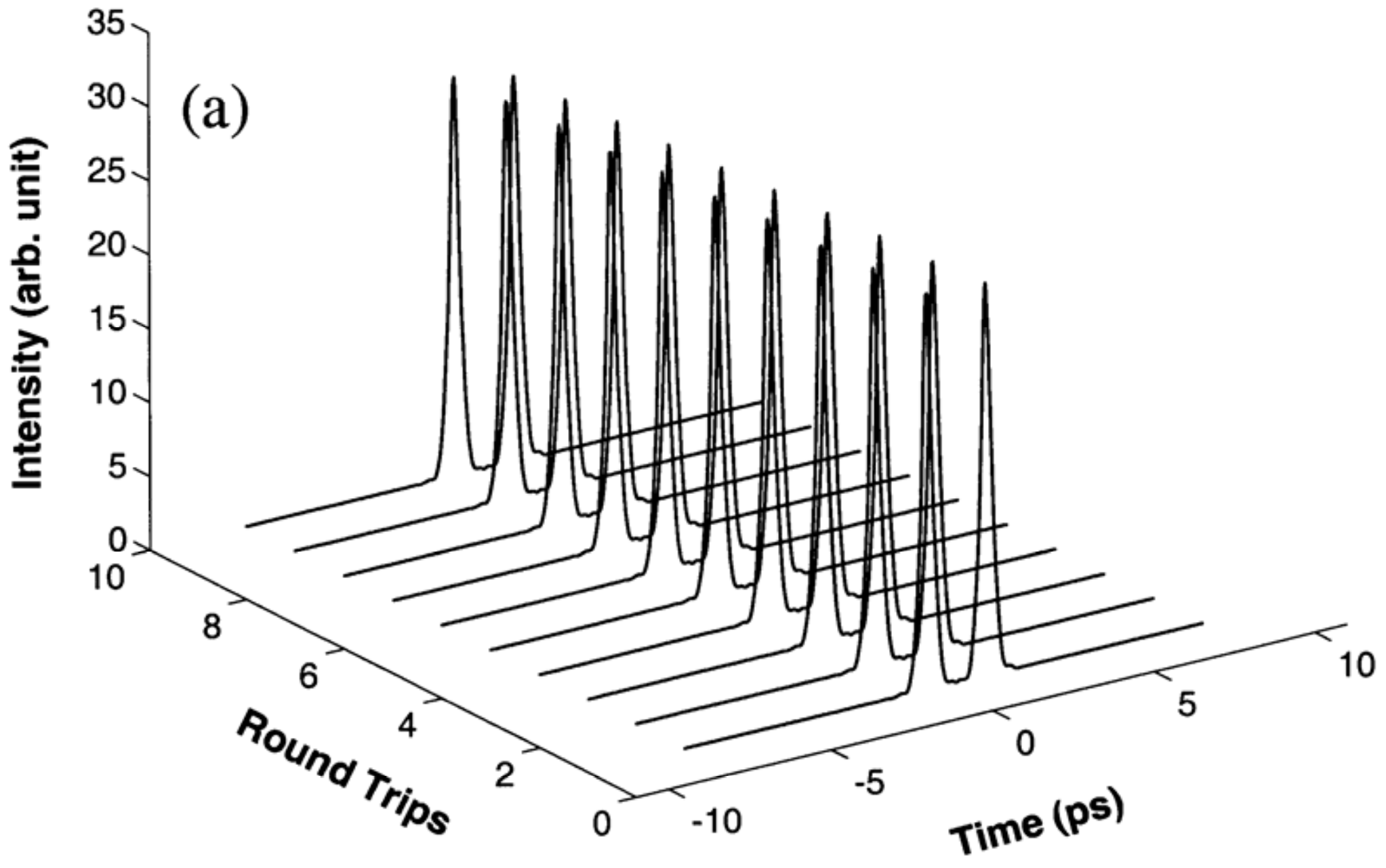

Figure 4(a).

D. Y. Tang *et. al.*

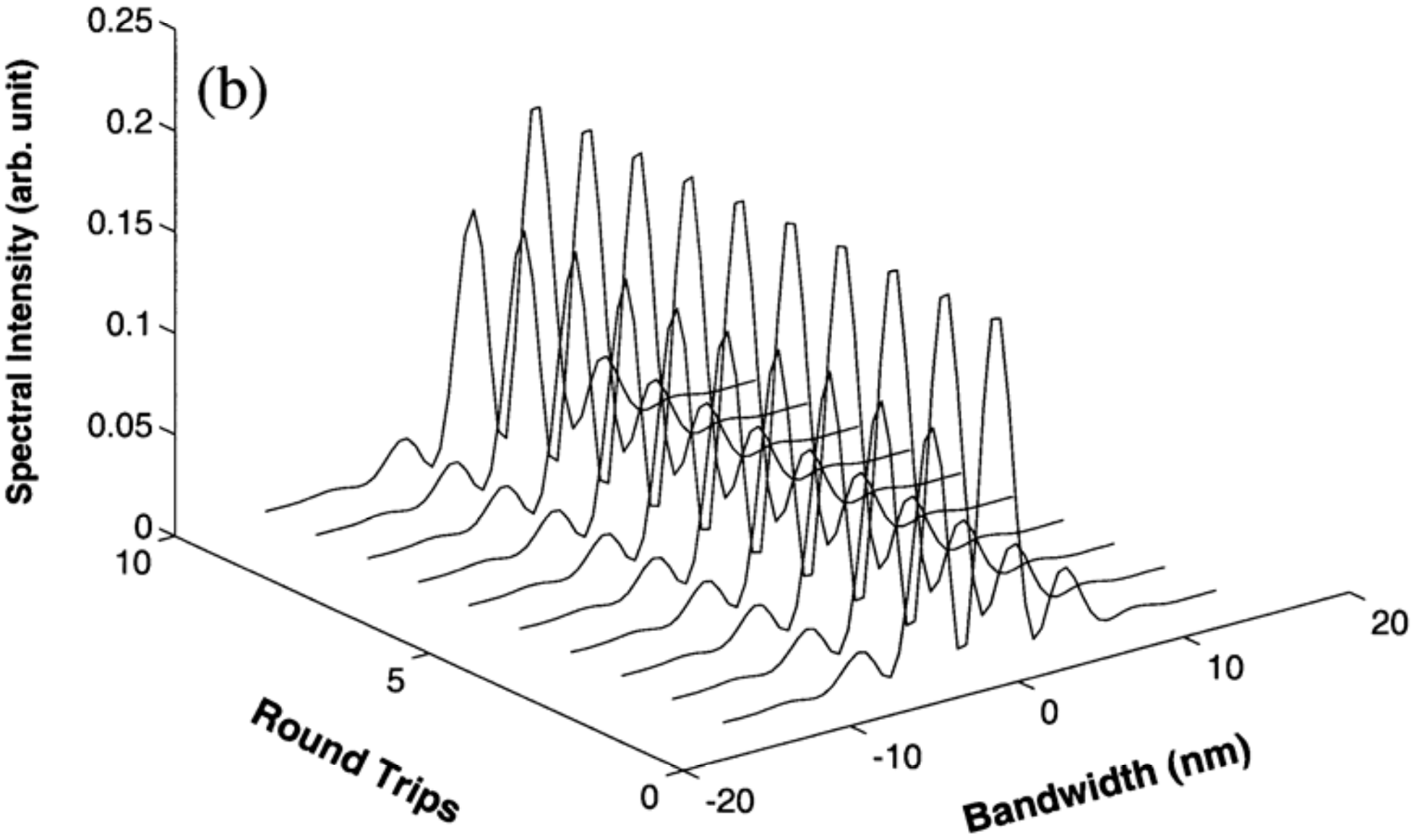

Figure 4(b).

D. Y. Tang *et. al.*